\documentclass[aps,prl,twocolumn,superscriptaddress,nofootinbib]{revtex4-1}
\usepackage[utf8]{inputenc}
\usepackage{graphicx}
\usepackage{xcolor}
\usepackage{amsmath}
\usepackage{ulem}
\begin{document}
\title{A microscopic picture of erosion and sedimentation processes in dense granular flows}
\author{Pierre Soulard}
\thanks{These two authors contributed equally.}
\affiliation{UMR CNRS Gulliver 7083, ESPCI Paris, PSL Research University, 75005 Paris, France.}
\author{Denis Dumont}
\thanks{These two authors contributed equally.}
\affiliation{Laboratoire Interfaces $\&$ Fluides Complexes, Universit\'e de Mons, 20 Place du Parc, B-7000 Mons, Belgium.}
\author{Thomas Salez}
\affiliation{Univ. Bordeaux, CNRS, LOMA, UMR 5798, F-33405 Talence, France.}
\affiliation{Global Station for Soft Matter, Global Institution for Collaborative Research and Education, Hokkaido University, Sapporo, Japan.}
\author{Elie Rapha\"el}
\affiliation{UMR CNRS Gulliver 7083, ESPCI Paris, PSL Research University, 75005 Paris, France.}
\author{Pascal Damman}
\email{pascal.damman@umons.ac.be}
\affiliation{Laboratoire Interfaces $\&$ Fluides Complexes, Universit\'e de Mons, 20 Place du Parc, B-7000 Mons, Belgium.}
\date{\today}

\begin{abstract}
Gravity-driven flows of granular matter are involved in a wide variety of situations, ranging from industrial processes to geophysical phenomena, such as avalanches or landslides. These flows are characterized by the coexistence of solid and fluid phases, whose stability is directly related to the erosion and sedimentation occurring at the solid-fluid interface. To describe these mechanisms, we build a microscopic model involving friction, geometry, and a nonlocal cooperativity emerging from the propagation of collisions. This new picture enables us to obtain a detailed description of the exchanges between the fluid and solid phases. The model predicts a phase diagram including erosion, sedimentation, and stationary-flow regimes, in quantitative agreement with experiments and discrete-element-method simulations.
\end{abstract}

\maketitle

\begin{figure}
\begin{center}
\includegraphics[width=\linewidth]{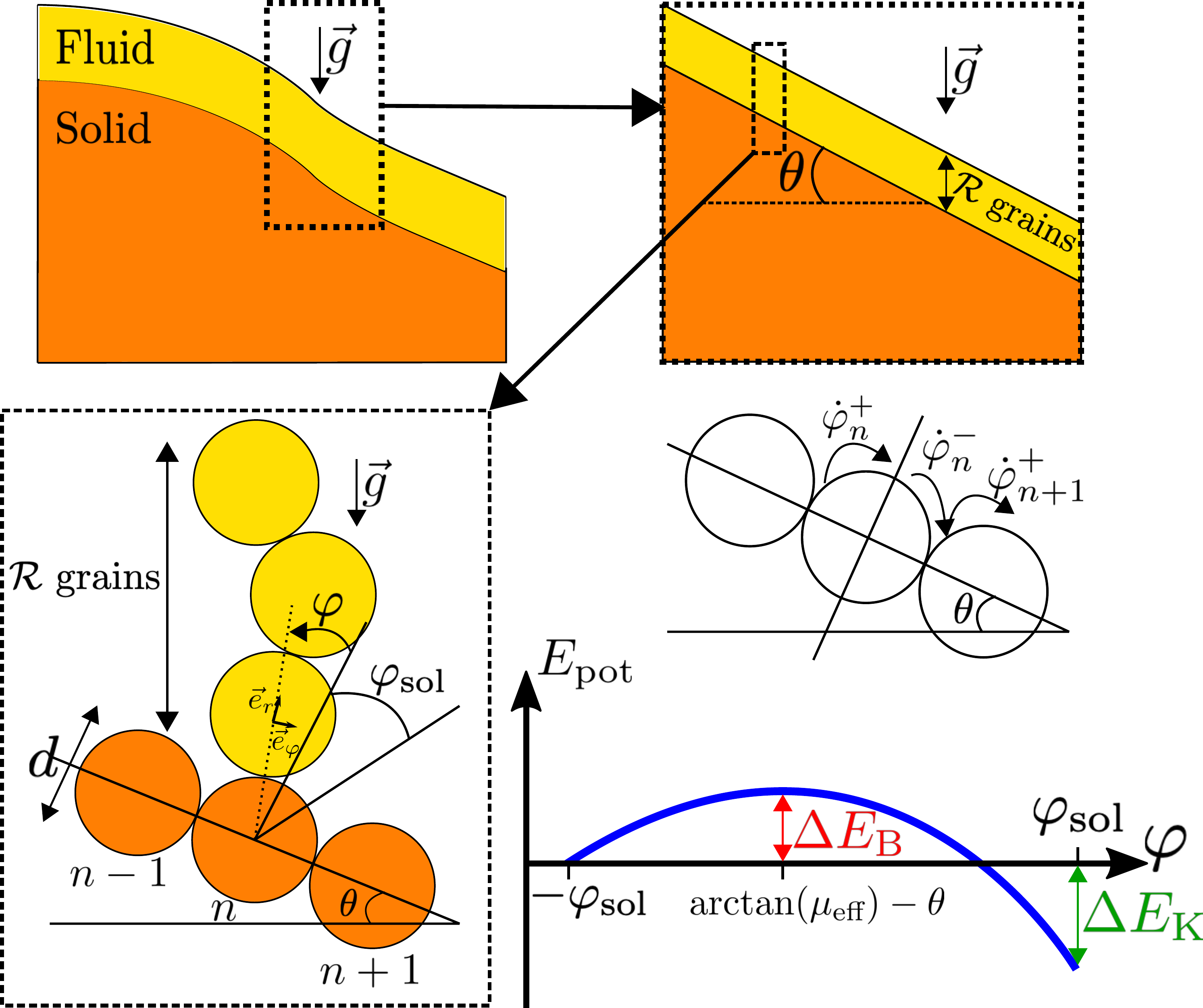}
\caption{A dense fluid granular phase (yellow) moves over an underlying static solid granular phase (orange). Locally, the solid-fluid interface makes an angle $\theta$ with the horizontal plane. The instantaneous position of a particular grain, moving above a static grain $n$, is described through the angle $\varphi$ between the normal to the interface and the contact interparticle-distance unit vector $\vec{e}_r$. The moving grain starts at angular position $-\varphi_\textrm{sol}$, with angular velocity $\dot{\varphi}_n^+$, and arrives at $+\varphi_\textrm{sol}$, with angular velocity $\dot{\varphi}_n^-$. Also shown is the effective potential energy $E_{\textrm{pot}}$ as a function of angular position $\varphi$, during the first step of the grain motion, within one cycle where $\varphi\in[-\varphi_{\textrm{sol}},+\varphi_{\textrm{sol}}]$. The two essential features are the barrier energy $\Delta E_{\textrm{B}}$ at maximum, and the kinetic energy gained at the end of the first step $\Delta E_{\textrm{K}}$.}
\label{fig:scheme}
\end{center}
\end{figure}

Over the last two decades, different theoretical approaches have been proposed to describe dense granular flows. The most widely used models are based on the $\mu(I)$ rheology~\cite{Jop2006}. This approach consists in a semi-empirical description of granular matter, through an effective friction coefficient $\mu$, that is a function of an inertial number $I$ directly related to the flow velocity~\cite{Andreotti2013}. 
Extensions of this $\mu(I)$ rheology have been proposed by several authors to address spatial heterogeneities and non-local effects~\cite{Pouliquen2009,Kamrin2012,Bouzid2013,Bouzid2015,Kamrin2019}. For instance, the yield stress increases as the thickness of the flow region reduces, which manifests itself through a thickness-dependent stop angle in granular flows down inclined planes~\cite{Pouliquen1999}. Another example deals with heap flows, where creep appears through a spatial exponential decay of the velocity over a few grains~\cite{Kamrin2019}. In parallel, semi-empirical models have been developed to describe the dynamics of avalanches, defined as dense granular flows atop static granular solids. This type of systems exhibits complex behaviours due to the intrinsic transfers of energy and matter between the fluid and solid phases at their interface. Specifically, the erosion of the solid phase by an avalanche feeds the fluid phase, and thus the avalanche, while the sedimentation of the fluid phase tends to stop the motion. The BCRE model was proposed to account for the coupled dynamics~\cite{BCRE1994,BCRE1995,BRdG1998,Trinh2017}. Two key ingredients are at the heart of this approach: i) the intuitive idea that the evolution of the interface between the two phases is determined by its local tilt angle $\theta$ (see Fig.~\ref{fig:scheme}); and ii) the assumed existence of a neutral angle $\theta^*$, such that for $\theta>\theta^*$ erosion occurs while for $\theta<\theta^*$ sedimentation occurs. Subsequent studies further suggested that the neutral angle is given by $\theta^*=\arctan (\mu_{\textrm{dyn}})$~\cite{Aradian2002}, where $\mu_{\textrm{dyn}}$ is an effective friction coefficient that might depend on the flow rate and the fluid layer thickness~\cite{Douady1999}. 

In this Letter, we propose a new microscopic description of the erosion and sedimentation processes at play in dense granular flows driven by gravity. Our model involves a flowing layer of grains over a static, yet erodible, one and includes a nonlocal cooperativity emerging from the propagation of collisions. Its predictions are directly confronted to numerical results obtained from discrete-element-method (DEM) simulations, in two canonical configurations: an inclined plane and a heap. Despite its simplicity, the proposed model enables us to obtain a detailed description of the exchange mechanisms between the fluid and solid phases, and ultimately a complete phase diagram of erosion and sedimentation. The model quantitatively describes the observed transitions between sedimentation, stationary flow, and erosion. Moreover, it allows to rationalize an important observable from inclined-plane experiments in the literature: the stop angle of a granular flow.

We consider a system made of identical spheres of diameter $d$, and mass $m$, with an interparticle sliding (resp. rolling) friction coefficient $\mu_\textrm{S}$ (resp. $\mu_\textrm{R}$)~\cite{Estrada2011}. These two coefficients are merged into a single effective coefficient $\mu_\textrm{eff}$ (see Supplementary Materials). 
It should be noted that our description is a mean-field one, involving average quantities and avoiding the actual irregularities inevitably present in a real granular medium.
As schematized in Fig.~\ref{fig:scheme}, we consider a moving layer of thickness $\mathcal{R}$ (in number of grains and counted vertically), \textit{i.e.} the fluid phase, above a static layer, \textit{i.e.} the solid phase. The roughness of the solid-fluid interface is characterized by an angle $\varphi_\textrm{sol}$ with respect to the normal to the interface, quantifying the angular depth of the hole between two grains. The value of $\varphi_\textrm{sol}$ ranges between $23.4^{\circ}$ and $30^{\circ}$ for spherical grains in 3D~\cite{Albert1997}. 
In the following, we focus on the motion of a single moving grain at the solid-fluid interface. This grain is subjected to the weight exerted by the $\mathcal{R}$ grains constituting the moving layer and is not allowed to jump. In the $\mu(I)$-rheology description, this situation is equivalent to considering flows with small inertial numbers. In order to move forward, the grain has to frictionally slide and/or roll over the bumpy underlying static layer. The instantaneous position of the moving grain is described through a single variable: the angle $\varphi$ between the normal to the interface and the contact interparticle-distance unit vector $\vec{e}_r$ (see Fig.~\ref{fig:scheme}).
 
As in previous studies~\cite{Quartier2000,Andreotti2001,Andreotti2007}, the motion is divided into two subsequent steps. First, the grain starts at angular position $-\varphi_{\textrm{sol}}$ with angular velocity $\dot{\varphi}_n^+$, and then moves above a static grain indexed by $n$, until it arrives at $+\varphi_{\textrm{sol}}$ with angular velocity $\dot{\varphi}_n^-$. Secondly, it elastically collides the next static grain indexed by $n+1$, which induces a sharp change of its velocity (orientation and norm), as well as secondary elastic collisions within the fluid and solid phases. A new and similar cycle then starts for the moving grain, with an initial angular velocity $\dot{\varphi}_{n+1}^+$. 

During the first step, the trajectory of the grain is circular. The only force components which contribute to the variation of its kinetic energy are along the tangential unit vector $\vec{e}_\varphi$. These are: the transverse projection $\mathcal{R}mg\sin(\varphi+\theta)\vec{e}_{\varphi}$ of the weight, and the effective friction force $-\mu_\textrm{eff}\mathcal{R}mg\cos(\varphi+\theta)\vec{e}_{\varphi}$ generated by the normal projection of the weight. We implicitly assumed here that most of the frictional dissipation occurs through the contact of the considered moving grain and the underlying static layer, since all the grains of the moving layer have similar velocities.

The total force along $\vec{e}_\varphi$ can be formally recast as the derivative of an effective potential, $-(1/d)\textrm{d}E_{\textrm{pot}}/\textrm{d}{\varphi}$, where ${E_{\textrm{pot}}(\varphi)}={\mathcal{R}mgd} \{\cos(\varphi+\theta)-\cos(\varphi_{\textrm{sol}}-\theta)+ \mu_\textrm{eff} \left[\sin(\varphi+\theta)+\sin(\varphi_{\textrm{sol}}-\theta)\right]\}$, and where the origin of energies has been arbitrarily chosen at $\varphi=-\varphi_{\textrm{sol}}$ (see Fig.~\ref{fig:scheme}). At $\varphi =\arctan(\mu_\textrm{eff})-\theta$, this potential is maximal, resulting in an energy barrier, that the grain has to overcome:
\begin{equation}
\Delta E_{\textrm{B}}=\mathcal{R}mgd [\sqrt{1+ \mu_\textrm{eff}^{2}}-\cos(\varphi_{\textrm{sol}}-\theta)+\mu_\textrm{eff}\sin(\varphi_{\textrm{sol}}-\theta)]\ .
\label{barriere}
\end{equation}
If the moving grain has enough initial kinetic energy $md^2(\dot{\varphi}_n^+)^2/2$ to overcome the barrier, it gains at the end of the first step the kinetic energy:
\begin{align}
\label{Delta_EK}
\Delta E_{\textrm{K}}&={md^2 \over 2} \ \left[(\dot{\varphi}_n^-)^2 - (\dot{\varphi}_n^+)^2 \right] ,\\
&=2\mathcal{R}mgd  \left[\sin (\theta) -\mu_\textrm{eff}\cos (\theta)\right] \sin (\varphi_{\textrm{sol}})\ .
\label{gain}
\end{align}

Up to now, we have not considered the contacts between grains within each of the two layers, leading to the nonlocal cooperative effects observed for dense granular flows. In our model, these naturally come into play during the second step of motion. Indeed, after the primary elastic collision with the $(n+1)^{\textrm{th}}$ static grain (see Fig.~\ref{fig:scheme}), the velocity of the considered moving grain changes suddenly, and cascades of secondary elastic collisions occur within the fluid and solid phases. This process leads to a cooperative energy reallocation that forms the essence of non-locality. Specifically, the energy transferred by the moving grain to the static grain during the primary collision is assumed to be proportional to the incoming energy, and reads $a md^2(\dot{\varphi}_n^-)^2/2$, with $a$ a constant prefactor comprised between 0 and 1. After the primary collision, the energies of the moving grain and the static grain thus temporarily become $(1-a) md^2(\dot{\varphi}_n^-)^2/2$ and $a md^2(\dot{\varphi}_n^-)^2/2$, before energy reallocation. Then, cascades of secondary elastic collisions are triggered in both phases. In a minimal description, we assume that they involve: i) $\mathcal{N}_\textrm{flu}$ and $\mathcal{N}_\textrm{sol}$ grains in the fluid and solid phases, respectively; and ii) some energy equipartition among those grains. 

For the fluid phase in particular, the energy primarily lost by the moving grain is redistributed over the $\mathcal{N}_\textrm{flu}$ moving grains. Over the primary and secondary collisions, the total energy loss $\Delta E_0=md^2[(\dot{\varphi}_n^-)^2-(\dot{\varphi}_{n+1}^+)^2]/2$ for the moving grain thus reads:
\begin{equation}
\Delta E_0= {amd^2 \over 2\mathcal{N}_\textrm{flu}} (\dot{\varphi}_n^-)^2\ .
\label{deltaezero}
\end{equation}
As a consequence, the ratio $\alpha=(\dot{\varphi}^{+}_{n+1}/\dot{\varphi}^{-}_{n})^2$ between the kinetic energies after and before the collisions is given by $\alpha= 1 - a/\mathcal{N}_\textrm{flu}$, and thus represents a direct signature of cooperativity in the fluid phase. Invoking the previous relations and Eq.~(\ref{Delta_EK}), one finally gets the central recursive equation:
\begin{equation}
(\dot{\varphi}^{+}_{n+1})^2=\alpha 
 \left[(\dot{\varphi}_n^+)^2+\frac{2\Delta E_{\textrm{K}}}{md^2}\right]\ .
\label{Shock}
\end{equation}

Assuming a global translational invariance in the system, and thus looking for the fixed point of Eq.~(\ref{Shock}), one gets $(\dot{\varphi}_{\infty}^+)^2= 2\alpha \Delta E_{\textrm{K}}/[md^2(1-\alpha)]$. In this homogeneous state, during each cycle, the kinetic energy gained by the moving grain when getting down the effective potential is exactly compensated by the loss due to the subsequent collisional process, such that $\Delta E_{\textrm{K}}=\Delta E_0$. 
The homogeneous state is stable only if the associated kinetic energy $E_{\textrm{K}}^{\infty} = md^2 (\dot{\varphi}_{\infty}^+)^2/2 =\alpha\Delta E_{\textrm{K}}/(1-\alpha)$, set by Eq.~(\ref{gain}), is larger than the barrier $\Delta E_{\textrm{B}}$, set by Eq.~(\ref{barriere}), giving a limiting condition :
\begin{equation}
\label{eq:energy_balance1}
\frac{\alpha}{1-\alpha} \, \Delta E_{\textrm{K}}(\theta_{\textrm{sed}}, \mathcal{R})= \Delta E_{\textrm{B}}(\theta_{\textrm{sed}}, \mathcal{R})\ ,
\end{equation}
which determines the sedimentation angle $\theta_{\textrm{sed}}(\mathcal{R})$. If $\theta<\theta_{\textrm{sed}}$, the lowest layer (at least) of the fluid phase stops, and $\mathcal{R}$ decreases.  

A similar reasoning allows us to discuss erosion. Considering the solid phase now, the energy $a md^2(\dot{\varphi}_n^-)^2/2$ gained by the static grain at the solid-fluid interface after the primary collision is then redistributed over the $\mathcal{N}_{\textrm{sol}}$ static grains through the cascade of secondary elastic collisions. In the homogeneous state, and over one complete cycle of motion and collisions, the solid phase receives a global energy $a md^2(\dot{\varphi}_{\infty}^-)^2/2=\mathcal{N}_\textrm{flu}\Delta E_{\textrm{K}}$. Invoking the energy equipartition among the $\mathcal{N}_\textrm{sol}$ grains, the static grain at the solid-fluid interface thus receives an overall net kinetic energy $\mathcal{N}_\textrm{flu}\Delta E_{\textrm{K}}/\mathcal{N}_\textrm{sol}$, set by Eq.~(\ref{gain}). The homogeneous state is stable only if this kinetic energy remains smaller than the energy required to drive the static grain into motion, given by the energy barrier $\Delta E_{\textrm{B}}(\theta, \mathcal{R}+1)$, set by Eq.~(\ref{barriere}). The limiting condition:
\begin{equation}
\label{eq:energy_balance2}
\frac{\mathcal{N}_\textrm{flu}}{\mathcal{N}_\textrm{sol}}\Delta E_{\textrm{K}}(\theta_{\textrm{ero}}, \mathcal{R})=\Delta E_{\textrm{B}}(\theta_{\textrm{ero}}, \mathcal{R}+1)\ ,
\end{equation}
determines the erosion angle $\theta_{\textrm{ero}}(\mathcal{R})$. If $\theta>\theta_{\textrm{ero}}$, the highest layer (at least) of the solid phase starts to flow, and $\mathcal{R}$ increases. The right-hand-side term of Eq.~(\ref{eq:energy_balance2}) represents the absolute value of a cohesion energy per grain of the solid phase, at the solid-fluid interface. For the dry granular assembly considered here, only solid friction and geometry control the cohesion of the solid phase. However, other binding mechanisms may be implemented in the model, such as capillary forces for wet granular matter~\cite{Saingier2017}. In an extreme case, the static grains can even be glued, as in inclined-plane experiments~\cite{Pouliquen1999}.
 
To solve Eqs.~(\ref{eq:energy_balance1}) and~(\ref{eq:energy_balance2}), we need to specify further $\mathcal{N}_\textrm{sol}$ and $\mathcal{N}_\textrm{flu}$, and thus the shape and size of the cooperative regions in each phase. In the bulk, we assume a cooperative region of either phase to contain $\xi$ grains and to have a fractal dimension $D$. Therefore, its typical length scale is given by $\sim\xi^{1/D}$. When the thickness of a phase becomes smaller than $\xi^{1/D}$, the cooperative regions of this phase are altered by the boundaries and the bulk description should be modified. In a heap-flow configuration, the solid phase is deep enough to ensure that $\mathcal{N}_\textrm{sol}=\xi$. In contrast, for the fluid phase, the fluid-air interface acts as a free-volume reservoir which is expected to truncate the neighbouring cooperative regions. At small $\mathcal{R}/\xi^{1/D}$, the number of grains in the cooperative regions of the fluid phase thus becomes $\sim\xi^{1-1/D} \, \mathcal{R}$, while at large $\mathcal{R}/\xi^{1/D}$ it saturates to $\xi$. To interpolate these two limiting behaviours, we propose the following Ansatz: $\mathcal{N}_\textrm{flu}(\mathcal{R})=\xi \left[1-\exp \left(-\mathcal{R}/\xi^{1/D}\right)\right]$. It should be noted that the precise functional form defining $\mathcal{N}_\textrm{flu}(\mathcal{R})$ is not crucial, as other expressions produce similar results (see Supplementary Materials). The key point here is that $\mathcal{N}_\textrm{flu}$ first increases with $\mathcal{R}$ before saturating.

\begin{figure}
\begin{center}
\includegraphics[width=1\linewidth]{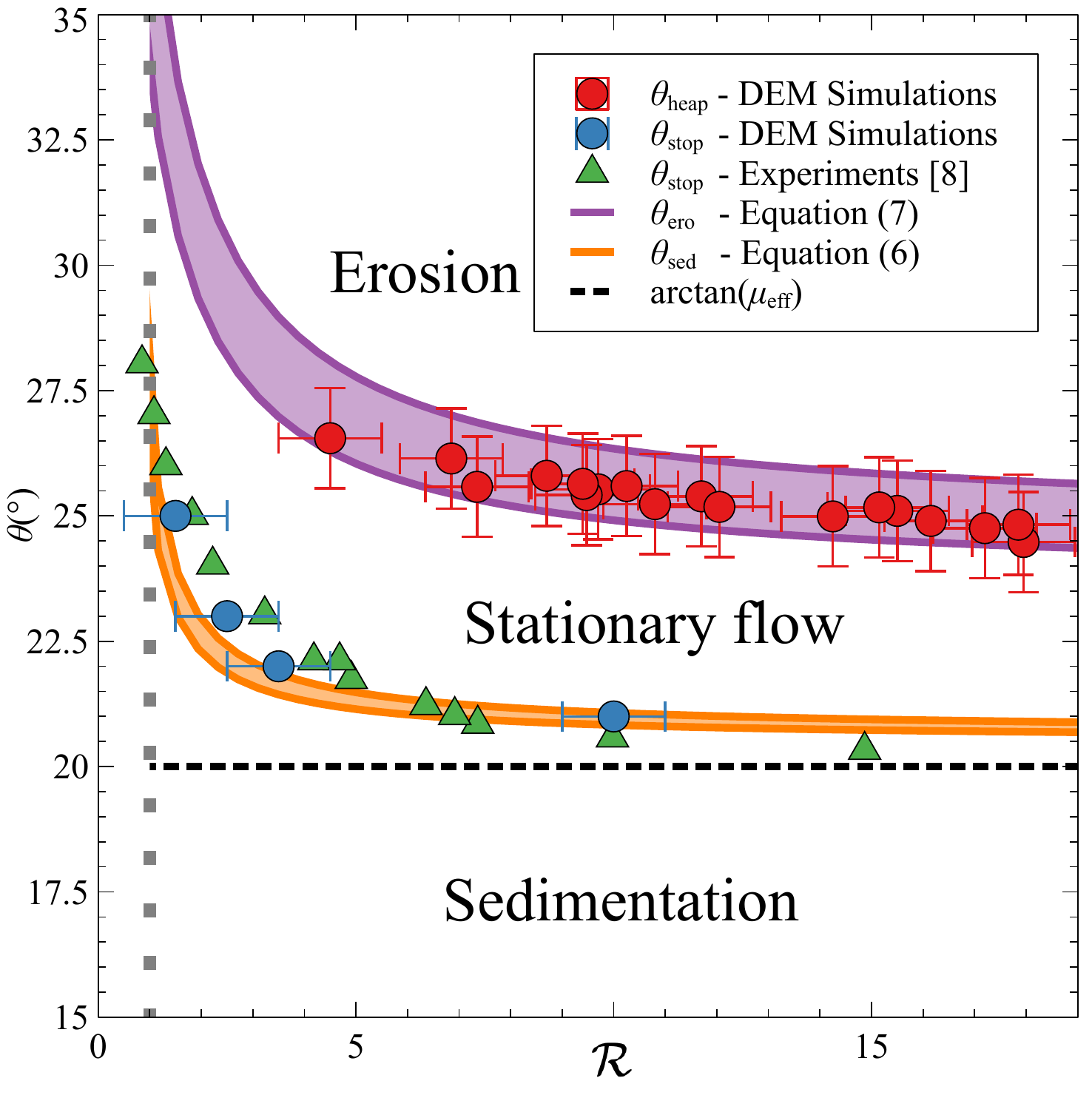}
\caption{The $\theta$-$\mathcal{R}$ phase diagram. Ensemble of results from: i) DEM simulations in two canonical configurations of granular flows, \textit{i.e.} inclined plane ($\theta_{\textrm{stop}}$) and heap ($\theta_{\textrm{heap}}$) ; ii) inclined-plane experimental results ($\theta_{\textrm{stop}}$)~\cite{Pouliquen1999} ; and iii) predictions ($\theta_{\textrm{ero}}$, $\theta_{\textrm{sed}}$) from the model through solutions of Eqs.~(\ref{eq:energy_balance1}) and~(\ref{eq:energy_balance2}). In the model, the fixed parameters are $\mu_\textrm{eff}=\tan(20^\circ)$~\cite{Pouliquen1999}, $a=0.5$~\cite{Quartier2000}, and $\varphi_{\textrm{sol}}\in[23.4^{\circ}, 30^{\circ}]$ (see light purple and light orange regions in the diagram)~\cite{Albert1997}, and the two adjustable parameters are $\xi=4.7$ and $D=0.94$. The horizontal dashed line indicates $\theta=\arctan(\mu_\textrm{eff})$, and the vertical one indicates $\mathcal{R}=1$.}\label{fig:recap}
\end{center}
\end{figure}

With the above expressions for $\mathcal{N}_\textrm{sol}$ and $\mathcal{N}_\textrm{flu}$, the evolutions of $\theta_{\textrm{sed}}$ and $\theta_{\textrm{ero}}$ can be computed by solving numerically Eqs.~(\ref{eq:energy_balance1}) and~(\ref{eq:energy_balance2}). There are five relevant dimensionless parameters in the model: $\mu_\textrm{eff}$, $\varphi_{\textrm{sol}}$, $a$, $\xi$, and $D$. The coefficient $a$ was estimated to be close to 0.5~\cite{Quartier2000}, while the effective friction coefficient is fixed to $\mu_\textrm{eff}=\tan(20^\circ)$, and $\varphi_{\textrm{sol}}$ spans the range $\-[23.4^{\circ}, 30^{\circ}]$~\cite{Albert1997}. Thus, $\xi$ and $D$ are the only free parameters. 
As shown in Fig.~\ref{fig:recap}, $\theta_{\textrm{sed}}$ and $\theta_{\textrm{ero}}$ rapidly decrease as $\mathcal{R}$ increases over a typical length scale $\sim\xi^{1/D}$, before saturating at two different values for thick fluid phases.
In the latter limit, the sedimentation angle appears to be close to $\arctan(\mu_\textrm{eff})$, while the erosion angle is clearly above. These two curves define sedimentation and erosion, and collectively determine the phase diagram for a flowing layer of grains atop a static one. Below the $\theta_\textrm{sed}(\mathcal{R})$ curve, the flow is unstable. Indeed, the moving grains at the solid-fluid interface do not have enough kinetic energy to overcome their effective potential barrier (see Fig.~\ref{fig:scheme}). Therefore, the thickness of the fluid phase decreases continuously and full sedimentation eventually occurs. Above the $\theta_\textrm{ero}(\mathcal{R})$ curve, the energy transmitted from the fluid phase to the solid one through collisions is sufficient for the barrier (\textit{i.e.} cohesion) energy of the static grains at the solid-fluid interface to be overcome. Therefore, the thickness of the fluid phase increases continuously and full erosion eventually occurs. In between these two curves, there is neither sedimentation nor erosion, and the flow can thus be considered as stable, \textit{i.e.} stationary.

To test the predictions of the model, one should ideally investigate the flow stability for experiments or numerical simulations with various ($\theta, \mathcal{R}$) sampling the three domains of the phase diagram. Unfortunately, for practical reasons, it is difficult to design such tests. Indeed, finite-size heap-flow experiments involve a continuous injection of grains at the top and their removal at the bottom. In this context, the fluid-phase thickness $\mathcal{R}(Q)$ and heap angle $\theta_{\textrm{heap}}(Q)$ self-adjust to the incoming flux $Q$ of grains, until a stationary state is reached -- but they cannot be controlled independently. In practice, one thus only has access to a unique relation $\theta_{\textrm{heap}}(\mathcal{R})$. Nevertheless, by building the initial heap with a sufficiently large angle, \textit{e.g.} near $90^{\circ}$, one can enforce the system to start far up in the erosion domain. Subsequently, erosion occurs freely until the stationary zone is reached. The system is then expected to spontaneously stabilise at the upper limit of the stationary zone, \textit{i.e.} at the erosion limit, thus allowing for a practical determination of $\theta_\textrm{ero}(\mathcal{R})$. 

In order to probe the sedimentation limit, the inclined-plane configuration~\cite{Pouliquen1999,Silbert2001,Borzsonyi2008,Malloggi2015,Kamrin2015} is particularly relevant. Therein, the solid phase is made of a single layer of grains glued on the incline, and is thus not erodible ({\it i.e.} the right-hand-side cohesion energy in Eq.~(\ref{eq:energy_balance2}) becomes infinite). The central observable is the stop angle $\theta_{\textrm{stop}}$, defined as the minimal tilt angle of the substrate for which a stationary flow is observed. As discussed above for the model, when the angle of the solid-fluid interface is below $\theta_\textrm{sed}(\mathcal{R})$, the lowest layer of the fluid phase stops and $\mathcal{R}$ reduces by one unit. Based on the homogeneous picture provided by Fig.~\ref{fig:recap}, and in particular the monotony of $\theta_\textrm{sed}(\mathcal{R})$, the system is then expected to remain in the sedimentation domain. The sedimentation front thus propagates upwards in the granular assembly, until the whole system is stopped. From this qualitative picture, the stop angle should be identified to the sedimentation one, thus allowing for a practical determination of $\theta_\textrm{sed}(\mathcal{R})$.

\begin{figure}
\begin{center}
\includegraphics[width=\linewidth]{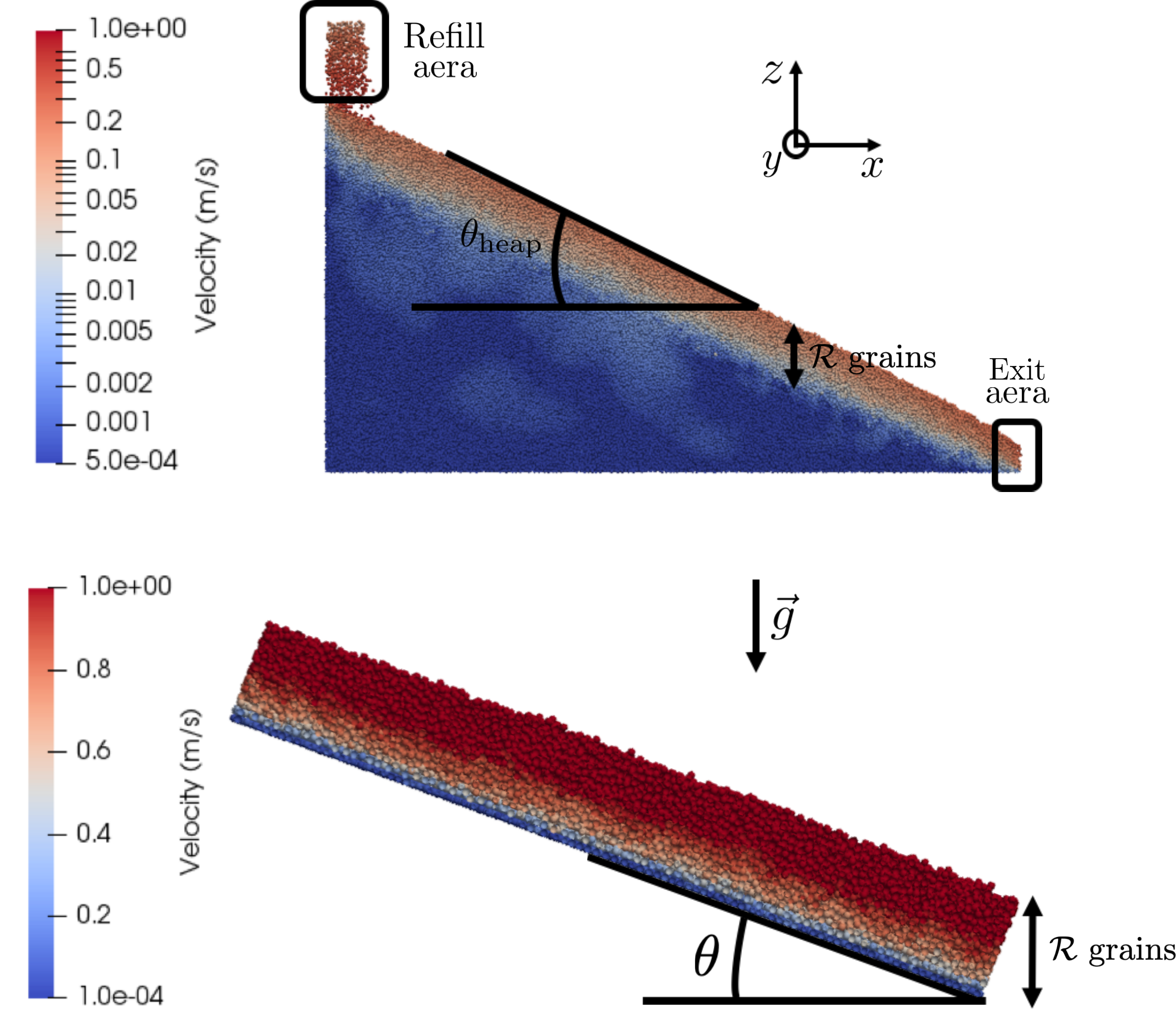}
\caption{Snapshots of the DEM simulations for two canonical configurations of granular flows: (top) heap flow with fixed flow rate; (bottom) inclined-plane flow with fixed angle and thickness. The colour maps refer to the velocity of the grains, as indicated in the scale bar.}
\label{fig:DEM}
\end{center}
\end{figure}

According to the previous considerations, we performed DEM numerical simulations for the two canonical experimental configurations: heap and inclined plane (see Fig.~\ref{fig:DEM}). The same microscopic parameters, \textit{i.e.} the same grains and friction coefficient, were used (see Supplementary Materials for technical details). As shown in Fig.~\ref{fig:recap}, the DEM numerical results for the two different configurations, together with the previously-reported experimental results for the inclined-plane configuration~\cite{Pouliquen1999}, are in quantitative agreement with the model predictions for identical parameters, under the proposed identifications: $\theta_{\textrm{heap}}\simeq \theta_{\textrm{ero}}$ and $\theta_{\textrm{stop}}\simeq \theta_{\textrm{sed}}$. The best-fit values of the two free parameters are $D=0.94$ and $\xi=4.7$, which suggest chain-like cooperative regions of a few grains and might be related to the force-chain network in static granular contact~\cite{geng2001,Majmudar2005}.    

The model having been corroborated by experiments and numerical simulations, we now turn to a detailed discussion of the phase diagram in Fig.~\ref{fig:recap}. First, sedimentation and erosion appear to be different processes, characterized by distinct angles, such that $\theta_\textrm{sed}(\mathcal{R})<\theta_\textrm{ero}(\mathcal{R})$. Therefore, stationary flows may exist for a continuum of angles ranging between these two disjoined boundaries. Interestingly, this opens a gap within the BCRE picture~\cite{BCRE1995}, in which a single neutral angle $\theta^*$ accounted for both the sedimentation and erosion processes. Besides, this may allow one to reconsider the previous suggestion of a friction coefficient depending on the fluid-layer thickness $\mathcal{R}$ -- that was made to rationalise the observations in the framework of a single neutral angle~\cite{Aradian2002,Douady1999}. Secondly, the validated decays of $\theta_\textrm{sed}$ and $\theta_\textrm{ero}$ with $\mathcal{R}$ are direct consequences of both the energy-equipartition and cooperativity-truncation ingredients in the model. Indeed, an increase of $\mathcal{R}$ implies that more grains of the fluid phase can share the energy losses due to the collisions with the static layer, which renders flow and erosion easier {\it i.e.} requiring smaller angles. Surface flow in yielded athermal granular media thus appears to be inhibited by the truncation of cooperativity -- as opposed to surface flow in ideal supercooled liquids at equilibrium~\cite{cst2015}. Thirdly, the saturations of $\theta_\textrm{sed}(\mathcal{R})$ and $\theta_\textrm{ero}(\mathcal{R})$ at large $\mathcal{R}$ are consistent with experimental observations showing that the flow properties of thick granular layers are independent of the fluid-phase thickness~\cite{Pouliquen1999,Andreotti2013}. 

In conclusion, we built a novel microscopic model involving friction, geometry, and nonlocal collisional effects, in order to describe erosion and sedimentation processes in dense granular flows atop static granular layers. As previously assumed~\cite{BCRE1994,BCRE1995,BRdG1998}, we have shown that erosion and sedimentation processes at the solid-fluid interface strongly depend on the interfacial angle. In contrast to previous work, our model suggests that each process might have its own critical angle, which depends on the fluid-layer thickness. The increase of both the erosion and sedimentation angles as the fluid-layer thickness decreases could be rationalised by considering two antagonist effects. Indeed, a decrease of $\mathcal{R}$ not only produces a lowering of the gravitational and thus frictional constraints for the moving grains at the solid-fluid interface, but induces as well a diminution of the size of the cooperative region in the fluid phase. This latter truncation mechanism impedes energy reallocation after collisional energetic losses, and dominates for small fluid-layer thicknesses. From our model, two classical experimental configurations -- inclined plane and heap -- usually analysed separately, could be described within a unified picture. A central outcome of our approach is a single phase diagram. Therein, the erosion and sedimentation limits were found to be in close agreement with numerical and past experimental measurements, of the heap and stop angles in inclined-plane and heap configurations respectively. These new results might be useful in the design of hydrodynamic models for dense granular flows, such as the BCRE scheme and depth-averaged method~\cite{Edwards2015}.

This work benefited from financial support by the Fonds National de la Recherche Scientifique (FNRS, PDR research project T.0109.16 ‘‘Capture biomim\'etique de fluide’’ and CDR project J.0191.17 ‘‘Mimicking elasticity with viscous fluids’’), and by the Action de Recherche Concert\'ee (UMONS, research project ‘‘Mecafood’’). D. D. acknowledges funding from FNRS through the Foundation for Training in Industrial and Agricultural Research. The authors also thank Yacine Amarouchene, Fabian Brau, and Paul Rambach for fruitful discussions.

\end{document}


\title{A microscopic picture of erosion and sedimentation processes in dense granular flows \\ -- Supplementary Materials --}
\author{Pierre Soulard}
\thanks{These two authors contributed equally.}
\affiliation{UMR CNRS Gulliver 7083, ESPCI Paris, PSL Research University, 75005 Paris, France.}
\author{Denis Dumont}
\thanks{These two authors contributed equally.}
\affiliation{Laboratoire Interfaces $\&$ Fluides Complexes, Universit\'e de Mons, 20 Place du Parc, B-7000 Mons, Belgium.}
\author{Thomas Salez}
\affiliation{Univ. Bordeaux, CNRS, LOMA, UMR 5798, F-33405 Talence, France.}
\affiliation{Global Station for Soft Matter, Global Institution for Collaborative Research and Education, Hokkaido University, Sapporo, Japan.}
\author{Elie Rapha\"el}
\affiliation{UMR CNRS Gulliver 7083, ESPCI Paris, PSL Research University, 75005 Paris, France.}
\author{Pascal Damman}
\email{pascal.damman@umons.ac.be}
\affiliation{Laboratoire Interfaces $\&$ Fluides Complexes, Universit\'e de Mons, 20 Place du Parc, B-7000 Mons, Belgium.}
\date{}
\maketitle

\section{Effective friction coefficient}
Let us first consider a solid object sliding on a solid plane while being submitted to a normal loading force $\mathcal{F}$ (with $\mathcal{F}>0$). The solid-solid friction at the interface is characterised by a sliding friction coefficient $\mu_\textrm{S}$. Over an infinitesimal tangential displacement $\textrm{d}x_\textrm{S}$ of the sliding object, the latter experiences an energy diminution $\delta W_\textrm{S}$ given by:
\begin{equation}\label{Sliding dissipation}
\delta W_\textrm{S} = \mu_\textrm{S}\mathcal{F}\lvert\textrm{d}x_\textrm{S}\rvert\ .
\end{equation}

Similarly, for a solid sphere of radius $a$ purely rolling on a plane while being submitted to a normal loading force $\mathcal{F}$ (with $\mathcal{F}>0$), the energy diminution $\delta W_\textrm{R}$ reads:
\begin{equation}\label{Rolling dissipation}
\delta W_\textrm{R} = \mu_\textrm{R}\mathcal{F}a\,\lvert\textrm{d}\Omega\rvert\ ,
\end{equation}
with $\mu_\textrm{R}$ the rolling friction coefficient and $\textrm{d}\Omega$ the infinitesimal angular variation.

Therefore, given the similarity between Eqs.~(\ref{Sliding dissipation}) and~(\ref{Rolling dissipation}),  it is reasonable to assume that when a spherical grain experiences both sliding and rolling friction the total energy diminution $\delta W$ is of the form:
\begin{equation}\label{Effective dissipation}
\delta W = \mu_\textrm{eff}\mathcal{F}\lvert\textrm{d}x\rvert\ ,
\end{equation}
where $\mu_\textrm{eff}$ is an effective friction coefficient which takes into account both sources of friction -- sliding and rolling -- and where $\textrm{d}x$ is the net infinitesimal tangential displacement. This energy dissipation corresponds to a Coulomb-like friction force $F_\textrm{eff}$, opposed to the motion, and satisfying:
\begin{equation}\label{Effective force}
\lvert F_\textrm{eff}\rvert = \mu_\textrm{eff}\mathcal{F}\ .
\end{equation}

The value of the effective friction coefficient depends on the respective amounts of rolling and sliding in the motion. For a purely sliding (resp. rolling) sphere, one has $\mu_\textrm{eff}=\mu_\textrm{S}$ (resp. $\mu_\textrm{eff}=\mu_\textrm{R}$). To the best of our knowledge, it is not possible to know \textit{a priori} the relation between  $\mu_\textrm{eff}$ and both $\mu_\textrm{S}$ and $\mu_\textrm{R}$. The effective friction coefficient is a coarse-grained parameter encompassing rugosity, surface chemistry, geometry and motion.

\section{Cooperative ansatz}
As explained in the main text, after the primary elastic collision of the considered moving grain with the next static grain indexed by $n+1$ (see Fig.~1 in the main text), the velocity of the moving grain changes suddenly, and cascades of secondary elastic collisions occur within the fluid and solid phases. These cascades involve cooperative regions of $\mathcal{N}_\textrm{flu}$ and $\mathcal{N}_\textrm{sol}$ grains in total, respectively. 

We assume that, in the bulk, any cooperative region of any of the two phases contains typically $\xi$ grains and has a typical fractal dimension $D$. The bulk cooperative length is thus given by $\sim\xi^{1/D}$. Furthermore, we assume that the fluid-air interface truncates the cooperative regions of the fluid phase for thin enough fluid layers, \textit{i.e.} at small $\mathcal{R}/\xi^{1/D}$. The number of grains in such a truncated cooperative region of the fluid phase thus becomes essentially $\sim\mathcal{R}\,\xi^{1-1/D}$, while at large $\mathcal{R}/\xi^{1/D}$ it should saturate to the bulk value $\xi$. 

\begin{figure}
\begin{center}
\includegraphics[width=0.85\linewidth]{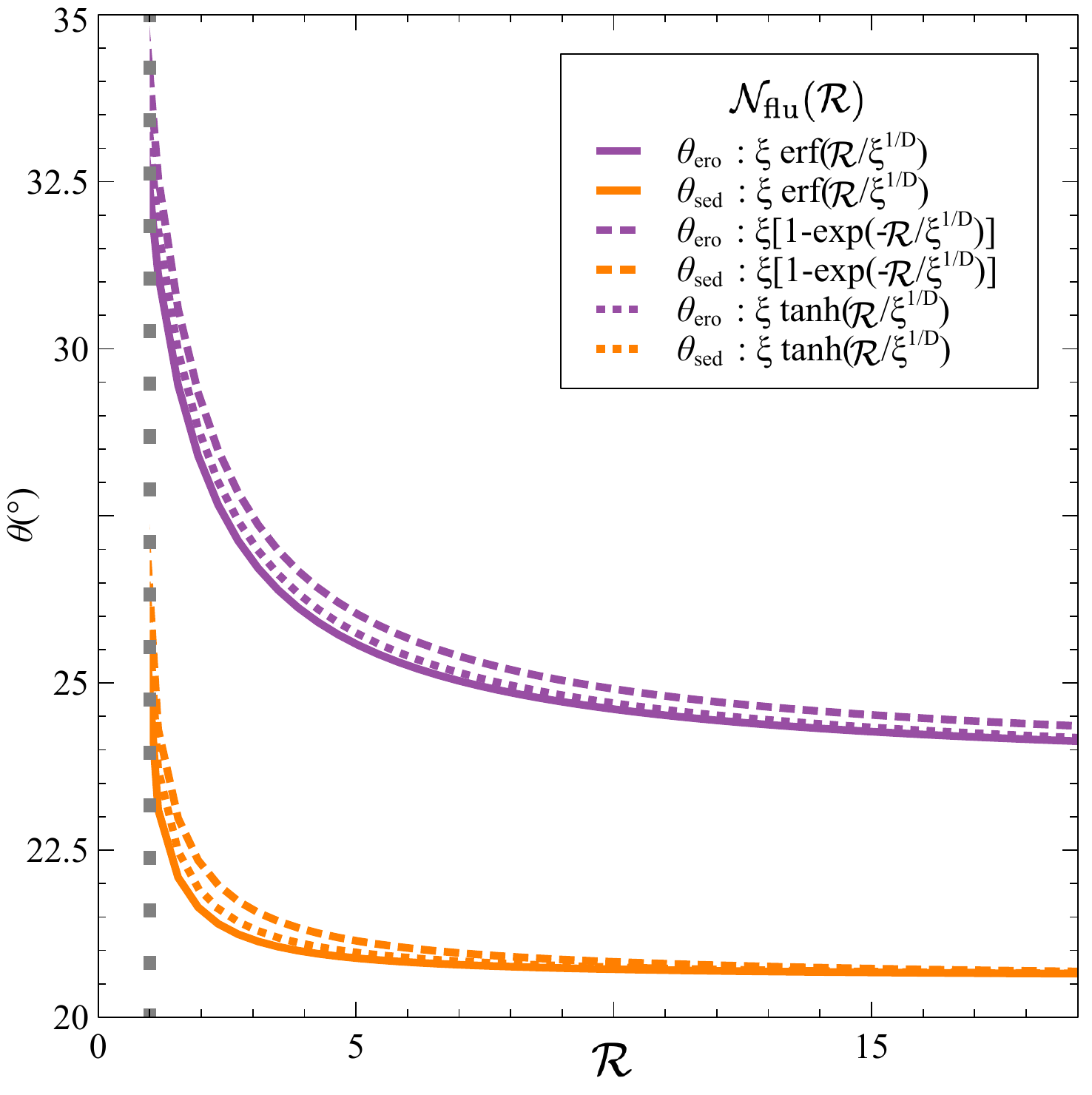}
\caption{Predictions of the sedimentation angle $\theta_\textrm{sed}$ and erosion angle $\theta_\textrm{ero}$ as functions of the fluid-layer thickness $\mathcal{R}$ (in grain-diameter unit), through solutions of Eqs.~(6) and~(7) in the main text, for three different functional forms for the $\mathcal{N}_\textrm{flu}(\mathcal{R})$ ansatz, as indicated. In the model, the parameters are $\mu_\textrm{eff}=\tan(20^\circ)$, $a=0.5$, $\varphi_{\textrm{sol}}=23.4^{\circ}$, $\xi=4.7$ and $D=0.94$ (see main text). The vertical dashed line indicates $\mathcal{R}=1$.}
\label{FigS1}
\end{center}
\end{figure}

To interpolate these two limiting behaviours, we have chosen the arbitrary ansatz: $\mathcal{N}_\textrm{flu}(\mathcal{R})=\xi \left[1-\exp \left(-\mathcal{R}/\xi^{1/D}\right)\right]$. However, Fig.~\ref{FigS1} shows that the exact mathematical form employed is not crucial, as other sufficiently sharp functions produce similar trends for $\theta_\textrm{sed}(\mathcal{R})$ and $\theta_\textrm{ero}(\mathcal{R})$. The only essential requirements are that $\mathcal{N}_\textrm{flu}(\mathcal{R})$ first increases with $\mathcal{R}$ before saturating to the bulk value.

\section{Numerical simulations}
Discrete Element Method (DEM) numerical simulations were performed with the software LIGGGHTS \cite{Kloss2012}. The simulated granular media were made of identical spherical beads with a $d=1$~mm diameter. The Hertz-Mindlin model was used to characterize the contacts between grains \cite{HertzMindlin}. The following micromechanical parameters were chosen in order to reproduce the macroscopic behaviour of realistic granular media: $0.5$ restitution coefficient, $\mu_{\textrm{S}}=0.5$ bead-bead sliding friction coefficient, $\mu_{\textrm{R}}=0.01$ bead-bead rolling friction coefficient, as well as $1$~MPa Young's modulus and $0.45$ Poisson ratio of the beads. 
In particular, both friction coefficients, $\mu_{\textrm{S}}$ and $\mu_{\textrm{R}}$, have been fixed in order to quantitatively reproduce the experimental results for spherical glass beads flowing down an inclined plane~\cite{Pouliquen1999}. The effective friction coefficient $\mu_{\textrm{eff}}$ of the granular medium is obtained by considering that $\mu_{\textrm{eff}} = \tan [\theta_{\textrm{stop}}(\mathcal{R}\rightarrow\infty)] \approx \tan(20^\circ)$.

For the inclined-plane configuration, we investigate the dependence of the stop angle with the fluid-phase thickness~\cite{Pouliquen1999,Silbert2001}. We use a rectangular channel of 100$d$ height, 100$d$ length and 20$d$ width, filled with a layer of beads of thickness $\mathcal{R}$ (in unit of $d$ and counted vertically). Periodic boundary conditions are applied along both length and width directions.
The plane is first inclined at an angle of 35$^\circ$, in order to set the layer into motion, and then the angle is rapidly fixed at a lower value $\theta$. After the system has reached a steady flowing state, the angle is then reduced again progressively until the flowing layer stops -- at the stop angle. In practice, the inclination is adjusted by artificially changing the direction of gravity.

\begin{figure}[h]
\begin{center}
\includegraphics[width=0.85\linewidth]{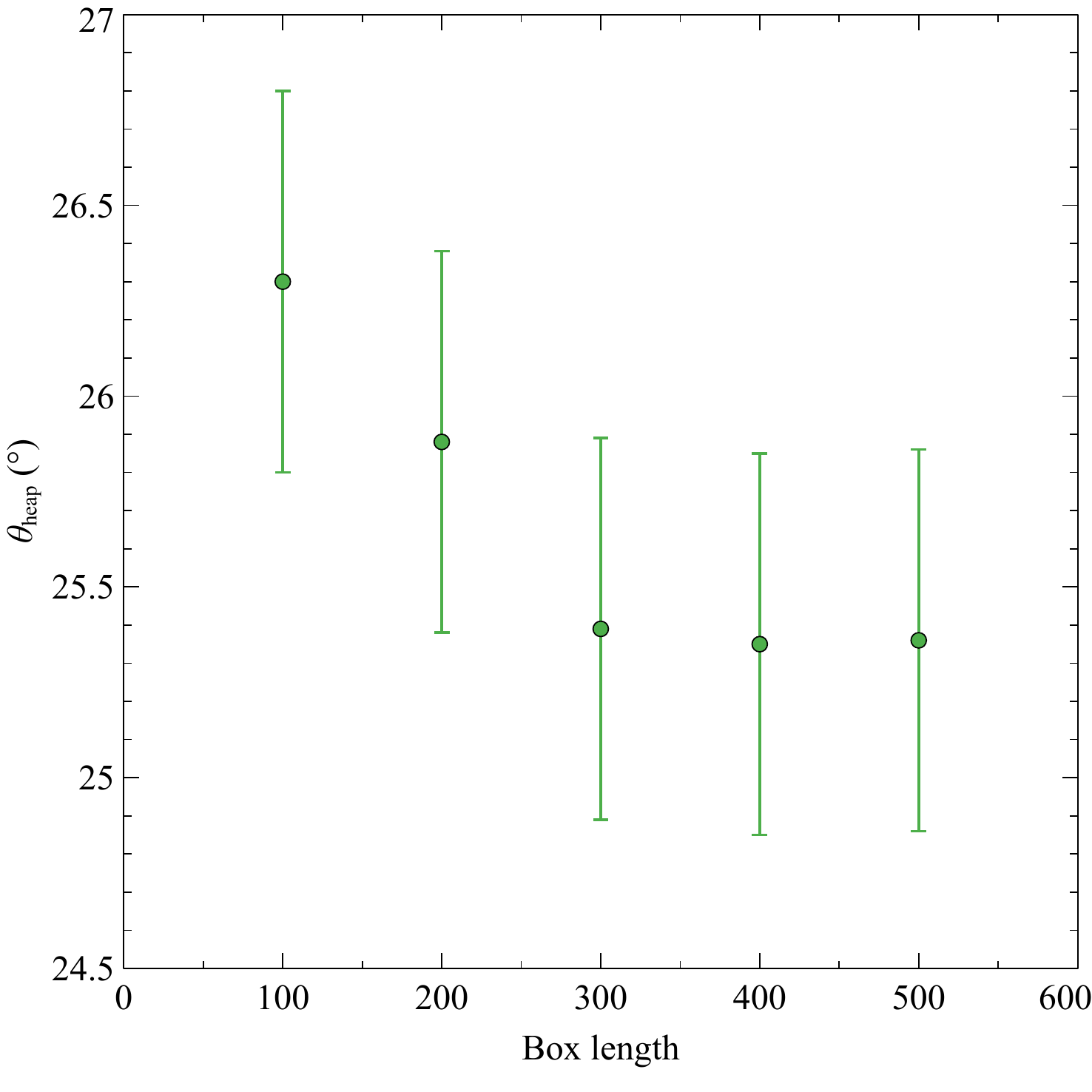}
\caption{Heap angle $\theta_{\textrm{heap}}$ as a function of simulation box length (in grain-diameter unit), from DEM simulations in the heap configuration with a flow rate $Q=21000$ grains/s.}
\label{FigS2}
\end{center}
\end{figure}

For the heap configuration, we quantify the fluid-phase thickness $\mathcal{R}$ and the angle $\theta_{\textrm{heap}}$ of the fluid-solid interface, in the steady state, for different externally-imposed flow rates $Q$. A rectangular box of 400$d$ height, 400$d$ length and 20$d$ width is initially filled with beads. Periodic boundary conditions along the width direction are applied. A slope appears due to the sudden removal of the wall at $x=400d$. A continuous refill starts at the top (see Fig.~3(b) in the main text) in order to compensate for the continuous loss of grains at the bottom exit. The simulation is ran until a steady state is reached. The system self-adjusts its thickness $\mathcal{R}$ and angle $\theta_{\textrm{heap}}$ for a given value of the flow rate $Q$. It should be emphasized that obtaining reliable measurements in DEM simulations can be difficult for the heap configuration. 
First, there is a drastic influence of lateral walls~\cite{Jop2005}, avoided here thanks to periodic lateral boundary conditions. 
Secondly, producing stationary flows down a heap requires very large systems. A large enough, 20$d$ wide, box is chosen in order to avoid any correlation due to the periodic boundary conditions. Moreover, as shown in Fig.~\ref{FigS2}, the length of the simulation box is also very important. Indeed, the observed heap angle $\theta_{\textrm{heap}}$ depends on the box length. For the systems studied here, we observed a saturation starting around a 300$d$ box length. Accordingly, we confidently chose a box with a 400$d$ length in order to avoid any effect of the box length. Thirdly, as previously suggested~\cite{Orpe2004,Orpe2007}, the self-adjusted fluid-layer thickness is only estimated through the position of the inflexion point in the velocity profile related to the solid-fluid crossover. Finally, as previously shown~\cite{Lemieux2000}, stationary flows cannot be obtained for the thinnest layers ($\mathcal{R} \lesssim 5$) in the heap configuration, as intermittent, unstable flows are instead observed.
